\documentstyle[aps,multicol,prb,epsf]{revtex}

\begin{document}
\ifpreprintsty \baselineskip=18pt \else\fi \bibliographystyle{simpl1}
\date{January 22, 1998}

\title{Spectral statistics in disordered metals: a trajectories
approach}

\author{R. A. Smith$^1$, I. V. Lerner$^1$, and B. L. Altshuler$^{2,3}$}

\address{$^1$School of Physics and Space Research, University of
Birmingham, Edgbaston, Birmingham~B15~2TT, United Kingdom\\
$^2$NEC Research Institute, 4 Independence Way, Princeton,
NJ 08540, USA\\
$^3$ Physics Department, Princeton University,  Princeton, 
NJ 08544, USA}

\maketitle
\begin{abstract}

We show that the perturbative expansion of the two-level correlation
function, $R(\omega)$,
 in disordered conductors can be understood semiclassically
in terms of self-intersecting particle trajectories. This requires the
extension of the standard diagonal approximation to include pairs of
paths which are non-identical but have almost identical action.
The number of diagrams thus produced is much smaller than in a  standard 
field-theoretical approach.  We show that such a simplification occurs
 because $R(\omega)$ has a natural representation as
 the second derivative of free energy $F(\omega)$.
 We calculate $R(\omega)$ to 3-loop
order, and verify a one-parameter scaling hypothesis for it in 2d.
We discuss the possibility of applying our ``weak diagonal approximation''
to generic chaotic systems.

\end{abstract}
\draft
\pacs{PACS numbers: 73.23.-b, 
73.20.Fz, 
05.45+b}  
\vspace*{-5mm}

\ifpreprintsty\else \begin{multicols}{2} \fi

The relationship between the quantum properties of disordered and
classically chaotic systems has been the focus of much recent
research\cite{ImSm,A2,A3,KhM2,A3S,A3S:NPh}. The main
reason for this is that both types of system show the same underlying
behavior -- their energy spectra in appropriate regions have
statistics given by random matrix theory (RMT)\cite{BGS,Ef:83}.
The spectral properties in these universal regions are well understood.
The challenge now is to understand whether generic features also emerge
in the deviations from universality. Answering this question is made
difficult because two  very different languages are 
used to characterize the two types of
system. In quantum disordered systems one averages over
all realizations of disorder to obtain an effective field theory: the
non-linear sigma model\cite{Ef:83,Weg:79,EfLKh}. In quantum chaotic
systems, (i.e. quantum systems which are chaotic in the classical limit),
one considers a particular system using the Gutzwiller trace 
formula\cite{Gut:71}, which involves a sum over classical periodic orbits.
Each language has its own strengths and weaknesses. The field theory
technique is rather formal and somewhat opaque physically, but has a
well-defined perturbation expansion. The trace formula appears to be
 more physically
transparent in that one is summing over classical trajectories. However,
 this sum is difficult to perform, its convergence properties are not well
understood, and no controlled expansion is currently available. It is
therefore natural that one should attempt to use the strengths of one
language to compensate the weaknesses of the other.

One recent attempt was to try to carry over the powerful calculational
techniques developed for the disordered systems to the chaotic 
systems\cite{KhM2,A3S}. In the most consistent form, one averages over
a certain energy interval to generate a field-theoretical functional
in the Wigner representation (the ``ballistic'' sigma-model). From this
one can derive, e.g., the two-level correlation function for quantum
chaotic systems. In the universal ergodic regime, the non-perturbative
derivation is equivalent to one which Efetov\cite{Ef:83} developed
for disordered systems, as
expected. In addition the leading-order perturbative term\cite{A3S} in the
non-ergodic regime gives a result equivalent to that obtained in the
diagonal approximation\cite{Berry} to the Gutzwiller trace 
formula\cite{Gut:71}. The same approximation when applied to the
disordered systems\cite{ImSm} also gives the leading order perturbative
term first derived in Ref.~\onlinecite{AS}. One therefore
expects that higher order perturbative terms in quantum
chaotic systems would be analogous to those occurring in disordered
systems (where they are known as weak-localization corrections). The
ballistic sigma model is, however, ill-defined in the ultraviolet limit,
making perturbative analysis currently ambiguous.

On the other hand, various authors\cite{LKh2,DK,ChSch,ImSm} have proceeded
in the opposite direction, and developed semiclassical methods for
the quantum disordered system. The goal is to gain a more intuitive
picture of phenomena such as weak localization and universal
conductance fluctuations.

In this paper we show how the diagrammatics for spectral correlations
in disordered conductors can be rewritten in terms of 
particle trajectories
which self-intersect in real space.  Each diagram for the two-level correlation
function is represented as consisting
 of two trajectories that are identical everywhere except
at self-intersection regions, where they are rejoined in different ways. 
As the two trajectories are identical for most of their length, they
are phase coherent, and interfere constructively. The perturbation
parameter, $1/g$, where $g$ is the the dimensionless conductance, can
be understood as the probability for a self-intersection to occur.
The great advantage of this approach is the drastic reduction in the
number of diagrams which occur in a given order of perturbation theory
This reduction
arises because our approach does not distinguish the starting points
of the (closed) trajectories, whereas the standard approach does. We
show that when translated into field theory language, this means that
the two-level correlation
function, $R(\omega)$, is the second derivative with respect to energy,
$\omega$, of the free energy, $F(\omega)$. As a particular problem we
calculate the two-level correlation
function to three-loop order in perturbation theory, which
yields the leading order level statistics for the unitary system in
2d. Finally we show that the three-loop results for $R(\omega)$
in both unitary
and orthogonal systems can be derived from a one-parameter scaling
hypothesis in which the renormalized conductance $g(\omega)$ is
substituted into the one-loop result. Note that in our approach we are
using the trajectory picture to classify the field theory diagrammatics
in contrast to the work discussed in the previous paragraph where
field theory is being used to classify periodic orbits.

We consider the relation of  the picture of trajectories 
to the perturbation expansion of $R(\omega)$ which is defined by
\begin{eqnarray}
\label{TLCF}
R(\omega)=\frac{1}{\nu^2}\Bigl<\nu(E+\omega)\nu(E)\Bigr> - 1\,.
\end{eqnarray}
Here $\nu(E)$ is the density of states per unit volume which can
be written, given the spectrum $\{E_n\}$ of the system, as
\begin{eqnarray}
\label{DoS}
\nu(E)=\frac{1}{L^d}\sum_n \delta(E-E_n)\,,
\end{eqnarray}
and $\nu=\left<\nu(E)\right>=1/(\Delta L^d)$, where $\Delta $ is the mean 
level spacing. Here $\langle\ldots\rangle$ denotes the averaging, either over
all realizations of disorder for a disordered system or over a certain energy
window for a chaotic system. Such energy windows are always narrow enough
so that $\nu$ has no energy dependence. The semiclassical approach allows one
to write  $\nu(E)$ as a sum over classical paths by means
of the Gutzwiller trace formula\cite{Gut:71},
\begin{eqnarray}
\label{Gutz}
\nu(E)=\sum_p A_p(E)\exp{\left(iS_p(E)/\hbar\right)}
\end{eqnarray}
where $A_p(E)$ and $S_p(E)$ are the amplitude and action of the $p$-th
periodic orbit at energy $E$. One can then substitute this semiclassical 
formula into the definition of the  two-level correlation
function, $R(\omega)$, and take the Fourier transform to
obtain the spectral form factor, $K(t)$,
\begin{eqnarray}
\label{SFF}
K(t)=\int \frac{d\omega}{2\pi\hbar}\exp{(-i\omega t/\hbar)}R(\omega).
\end{eqnarray}
Since all factors vary slowly with $E$ except the action $S_p(E)$,
one expands $S_p(E+\omega)$ to first order to obtain
\begin{eqnarray}
\label{SCSFF}
K(t)=\sum_{p,q}A_pA_q^*e^{i(S_p-S_q)/\hbar}
\delta[t-\frac{1}{2}(T_p+T_q)]
\end{eqnarray}
where $T_p(E)=dS_p(E)/dE$ is the period of the orbit $p$. 

To proceed further Berry\cite{Berry} 
introduced the diagonal approximation
assuming that only terms with $p=q$ contribute to the sum in
Eq.\ (\ref{SCSFF}).
The reasoning is that terms with $p\ne q$ have randomly varying phases and
cancel each other. In the disordered metal
such a diagonal approximation was shown \cite{ImSm} to
reproduce the leading order
perturbative contribution\cite{AS} to $R(\omega)$.

The question that we address here is how to get regular corrections 
to the diagonal approximation which are due to contributions of
relatively long trajectories. Such corrections are of fundamental importance,
especially for two-dimensional disordered systems: 
the leading order term corresponding to the  diagonal approximation
 vanishes\cite{KL:95a} for $d=2$. However, the possibility of getting  the 
 regular corrections could be interesting in a much wider context. To 
evaluate the corrections one has
 to  consider pairs of paths with {\it almost}  equal
actions, $S_p\approx S_q$,
like in the diagonal approximation,  but allow $p\ne q$.

To construct the {\it weak diagonal approximation} which 
allows for such pairs of
trajectories we start with  a path that self-intersects in real 
space at one or
more points. At a point of intersection we break up the path and make different
trajectories by joining the pieces  together in different
ways. We see at once that this gives rise to a perturbation expansion,
with the perturbative order given by the number of 
loops in real space created by self-intersections,
and the perturbation parameter given by the
probability of having a self-intersection. In the case of a disordered
electronic system, we will show that this expansion is none other
than the usual field-theoretical loop expansion. Before we do this, let us
speculate on the nature of this expansion for a generic chaotic system. 

In this case the above picture of self-intersecting trajectories
requires clarification. First of all, the picture of trajectories is formulated
in phase space rather than in 
real space. Of course, classical trajectories which
are identical along part of their length in phase space
cannot diverge in the way discussed above -- this process
is quantum mechanical in nature.  The uncertainty
principle means that phase space is coarse-grained  into boxes with size of
order $\hbar^d$. Two trajectories which were originally nearly identical
(i.e.\ passing through the same phase space boxes) start to 
deviate significantly at the Ehrenfest time, $t_{\text{Ehr}}$,
due to quantum processes such as diffraction. 
Then we can make pairs of piecewise identical trajectories as described above.
For this picture to make sense we 
 need to work at time-scales greater than  $t_{\text{Ehr}}$. This has been
noted in Ref.\ \onlinecite{AlLa1}
which recently considered the weak localization correction   
in a disordered system where the Ehrenfest time is determined by
the  diffraction at
randomly distributed scattering spheres.  In the usual model of a  disordered
conductor with point-like scatterers $t_{\text{Ehr}}$ 
 coincides with  the elastic scattering time $\tau$
since the particle's direction is totally randomized after each scattering.
 In general, we expect
that the classification of trajectories in the weak diagonal
approximation might be valid for chaotic system with $t_{\text{Ehr}}\ll
t_{\text{erg}}$ where $t_{\text{erg}}$ is the ergodic time scale at which the
trajectory samples the entire phase space.
This condition is necessary in order to treat the  self-intersection region
as a perturbation. It simply means that
 the length of  almost identical regions of the 
two trajectories is much 
larger than the length of the self-intersection region. 

At time scales $t\gg t_{\text{erg}}$, one expects completely universal behavior
which is described by the random matrix theory. \cite{RMT} 
This has been conjectured by Bohigas et al \cite{BGS}
and partially proved by Andreev et al. \cite{A3S,A3S:NPh}
This universal behavior is completely non-perturbative and cannot be
described by the diagonal approximation. 
Moreover, even for $t\ll t_H$, where $t_H\equiv \hbar /\Delta$ is the 
Heisenberg time, the diagonal approximation does not reproduce the deviation
from the linear behavior of $K(t)$ known from the random matrix theory.
We will show that we can exactly allow for this deviation within our
weak diagonal approximation. 

At time scales $t_{\text{Ehr}}\ll t\ll t_{\text{erg}}$,
one expects non-universal corrections to the diagonal approximation.
It is not clear whether this is true for a generic chaotic system. 
Attempts\cite{AgLS} to generate these corrections within the ballistic nonlinear
$\sigma $ model\cite{KhM2,A3S,A3S:NPh} have led to results vanishing to all
orders in perturbation theory. These results are dubious, however, because 
they depend strongly on the short-time regularization procedure. In particular,
there exist regularizations where the corrections do not vanish and are
totally governed by short-time processes at time scales $t\alt t_{\text{Ehr}}$. 
In addition, the ballistic $\sigma $ model itself might not take into account
all relevant quantum processes. 

In the disordered conductor these corrections can be calculated both for 
 $t\gg t_{\text{erg}}$ and for  $t\ll t_{\text{erg}}$. Here 
 $t_{\text{erg}}=L^2/D$ is the time taken for an electron to diffuse to the
boundaries. We will show that the trajectory picture is naturally related
to a standard diagrammatic expansion and, moreover, allows considerable 
simplifications in this expansion.

We will now draw all closed trajectories with up to
two self-intersections regions and show their relation to 
the diagrams up to three-loop order.
Let us first rewrite a pair of identical 
(up to discrete symmetries of the ensemble)  trajectories, which
correspond to  Berry's diagonal approximation, \cite{Berry,ImSm} 
as a diagram in field theory language.

\ifpreprintsty \else
\begin{figure}
\vspace{0.3cm}
\hspace{0.02\hsize}
\epsfxsize=0.9\hsize
\epsffile{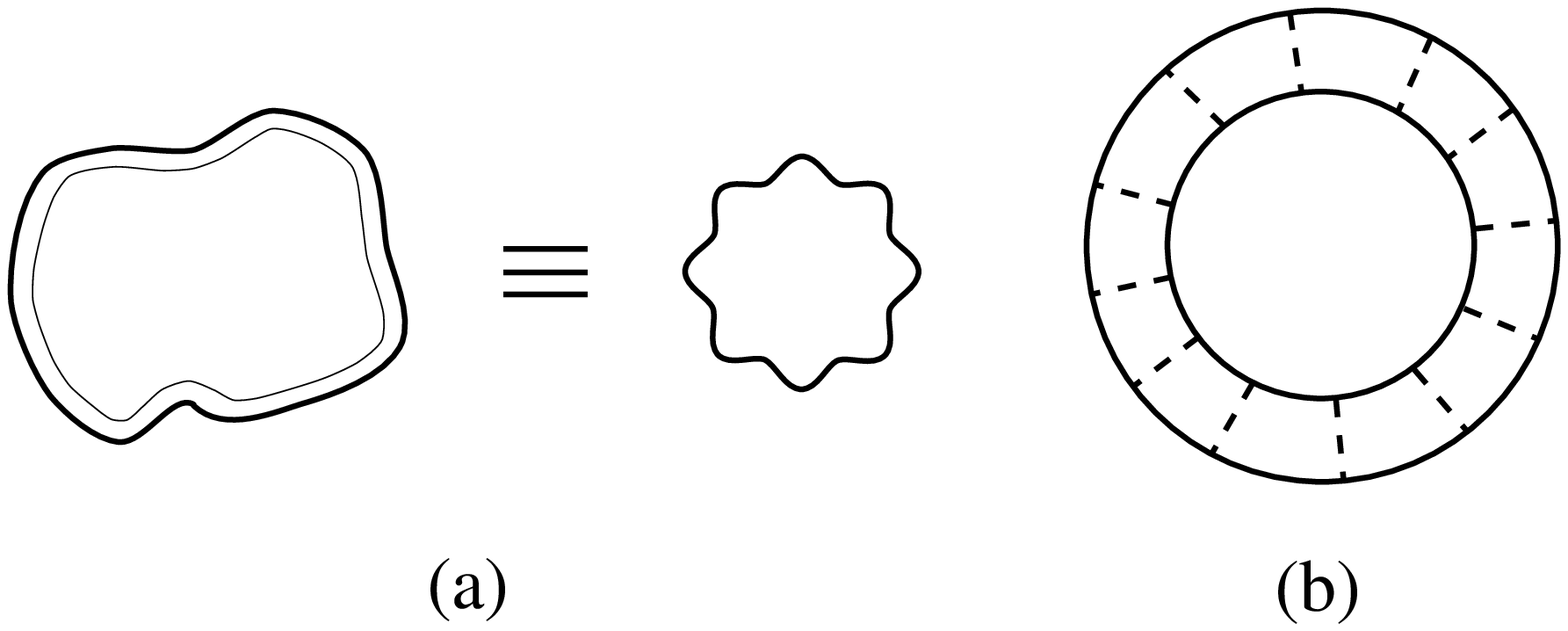}
\vspace{0.3cm}
 
\refstepcounter{figure}
\label{fig:1}
{\small \setlength{\baselineskip}{10pt} FIG.\ \ref{fig:1}.
(a) The 1-loop diagram for the free energy  $F(\omega)$, 
which consists of pairs of trajectories that are identical up to 
time reversal symmetry. In the field theory language it consists of 
a single closed wavy line. 
(b) The same diagram in the disordered metal. Since the closed wavy
line in (a) consists of a sum over impurity ladders, and the diagram
with $n$ impurity ladders has symmetry factor $1/n$, the ladder gives
logarithmic contribution $-\ln{(Dq^2-i\omega)}$.
 }
\end{figure}
\fi

\ifpreprintsty \else
\begin{figure}
\vspace{0.3cm}
\hspace{0.02\hsize}
\epsfxsize=0.9\hsize
\epsffile{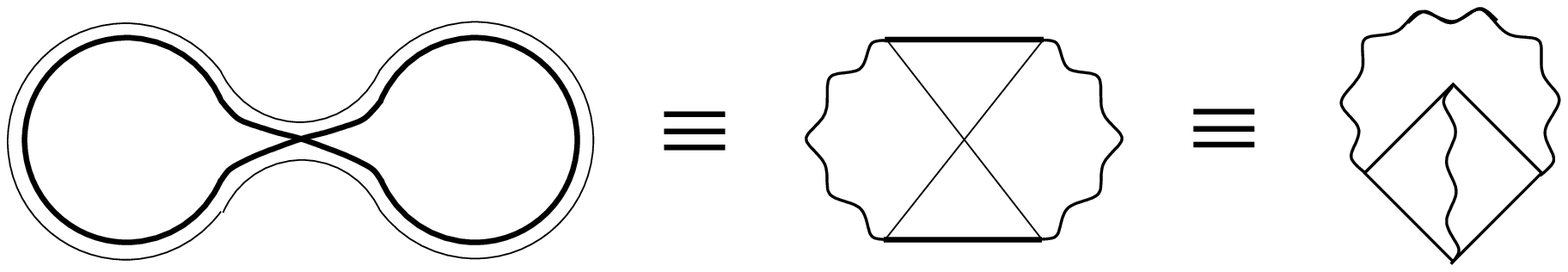}
\vspace{0.3cm}
 
\refstepcounter{figure}
\label{fig:2}
{\small \setlength{\baselineskip}{10pt} FIG.\ \ref{fig:2}.
The 2-loop diagram for the  free energy $F(\omega)$ 
corresponding to paths with one self-intersection. First we show 
the two different ways of linking up the paths at the 
self-intersection point. Then we rewrite this in the field theory 
language so that regions where the two paths are identical become 
wavy lines, whilst the region of self-intersection becomes a closed 
box. Finally we twist the latter diagram into its usual form. 
 }
\end{figure}
\fi

The pair of
identical  or time-reversed trajectories becomes a diffusion or Cooperon
propagator, respectively, drawn as a wavy line. This yields the one-loop
diagram shown in Fig.~1a. 

The two-loop case is shown in Fig.~2 and involves one self-intersection 
region.
At this point the paths can be linked up in two different ways, and the
diagram consists of these two trajectories. Again we rewrite in field theory 
language with the identical  or time-reversed
portions of path becoming diffusion or Cooperon propagators.
The region around the self-intersection becomes an effective
interaction between these propagators known as a Hikami box\cite{Hik:81}.
If we put arrows on the trajectories to show their direction of traversal
we see that they go in the same direction over part of the diagram. This
means that time-reversal invariance is required for their actions to be
identical. Hence there is no two-loop term in the unitary case, where
time-reversal invariance is broken; in fact, all even loop contributions
vanish for the same reason.

The three-loop case is shown in Fig.~3, and involves two 
self-intersections. The situation here is more complicated since there
are three distinct topological ways for two self-intersections to occur.
They can occur either at the same point, or at two different points
$A$ and $B$. In the latter case there are two distinct orderings in which
the trajectory can be traversed -- $AABB$ and $ABAB$. Finally there is
more than one way of linking up the partial paths to form trajectories.
There are five diagrams in total: $F_{3a}$ has $AABB$ form; $F_{3b}$ and
$F_{3c}$ have intersection at only one point; $F_{3d}$ and $F_{3e}$ have
$ABAB$ form. Putting arrows on the trajectories we find that only
$F_{3b}$ and $F_{3d}$ contribute to the unitary case.

We see that the above procedure can obviously be generalized to any order
of perturbation theory, and is a powerful way of ensuring that all
contributions have been considered.

Let us now relate these pictures of semiclassical trajectories to
the standard diagrammatic approach\cite{AS} for evaluating $R(\omega)$.
The starting point is the expression for the density of states,
Eq.\ (\ref{DoS}), in terms of electron Green's functions,
\ifpreprintsty \else\end{multicols} 
\fi
\vspace*{\fill}
\ifpreprintsty \else
\begin{figure}
\vspace{0.3cm}
\hspace{0.02\hsize}
\epsfxsize=0.9\hsize
\epsffile{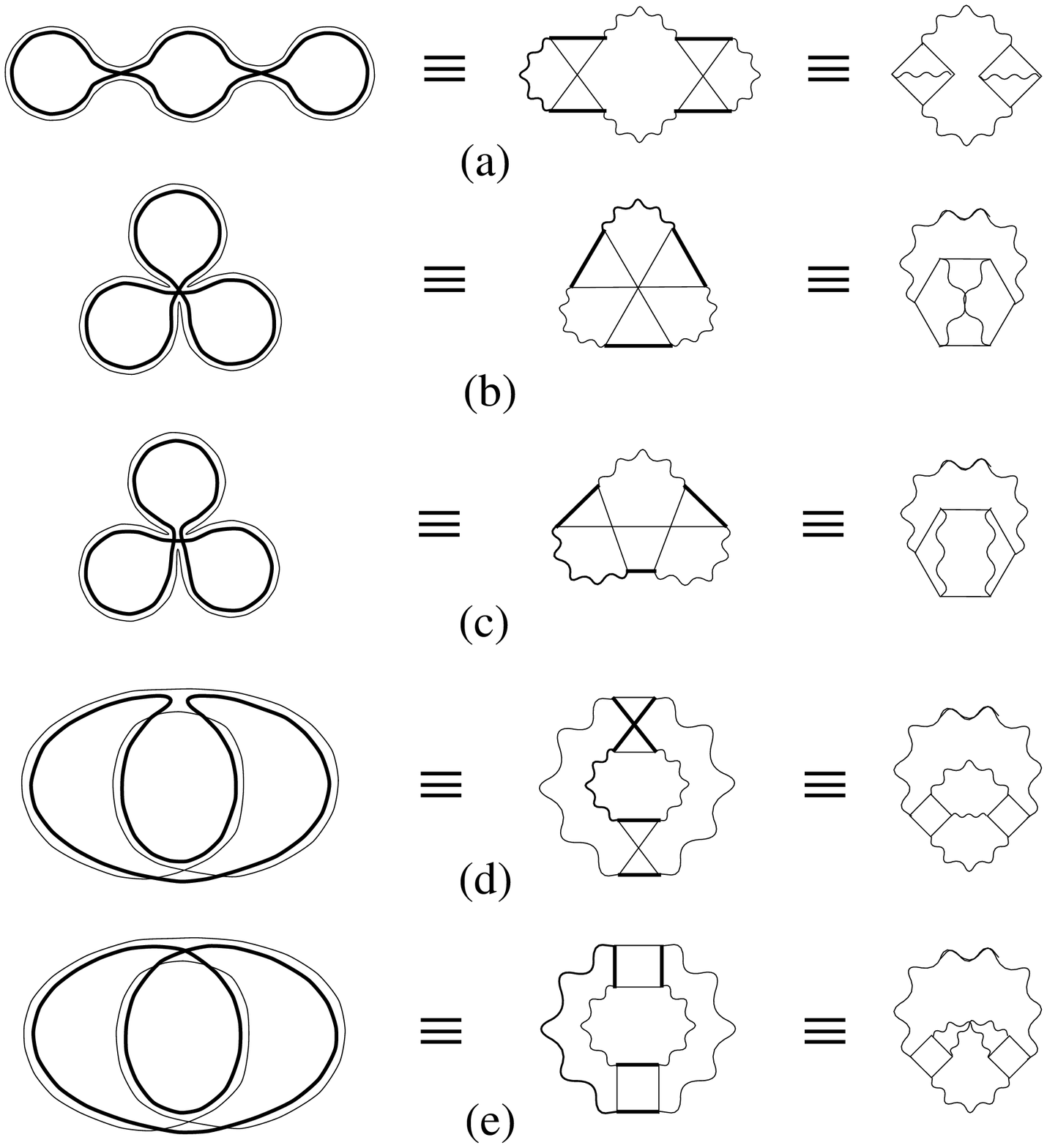}
\vspace{0.3cm}
 
\refstepcounter{figure}
\label{fig:3}\vspace{1cm}

{\small \setlength{\baselineskip}{10pt} FIG.\ \ref{fig:3}.
 The five 3-loop diagrams for the free energy $F(\omega)$ 
corresponding to paths with two self-intersections. We follow the 
format of Fig.~2, showing first the pairs of distinct trajectories; 
then rewriting these with wavy lines representing regions where the 
two paths are identical, and closed boxes at the self-intersection 
points; then finally twisting the latter diagrams in their usual form.
 }
\end{figure}
\fi
\vfill\break

\begin{eqnarray}
\label{GrGa}
\nu(E)=\frac{i}{2\pi L^d}\int \text{d}^dr \left[G^R(\bbox{r},\bbox{r};E)
-G^A(\bbox{r},\bbox{r};E)
\right]\,.
\end{eqnarray}
It follows from this expression and the definition (\ref{TLCF}) that
\begin{eqnarray}
\label{Reqn}
R(\omega)=\frac{\Delta^2}{2\pi^2}\Re \text{e}\!\int
\!\text{d}^dr \text{d}^dr'
\left < G^R(\bbox{r},\bbox{r};E_1)G^A(r',r';E_2)\right >_{\text{c}}\,,\qquad
\omega\equiv E_1-E_2\,.
\end{eqnarray}
	\ifpreprintsty \else 
                  	\vspace*{-3.5ex} 
			\begin{minipage}{0.48\hsize}$\,$\hrule
			 \end{minipage}\vspace*{1.5ex}
                  \begin{multicols}{2}
                   \fi
Here only the connected average of $ G^RG^A$ remains:
the sum of all unconnected averages is absorbed by $-1$ in the 
definition (\ref{TLCF}), while the connected average of $G^RG^R+G^AG^A$ 
vanishes.

Equation (\ref{Reqn}) directly corresponds to the semiclassical expressions
(\ref{SFF}) and (\ref{SCSFF}), since $ G^R(\bbox{r},\bbox{r};E)$
is the quantum mechanical
propagator for a particle of energy $E$ to start and finish at the same point
$r$. We can then represent $ G^R(\bbox{r},\bbox{r};E)$
as the sum of all closed 
paths while $ G^A(\bbox{r},\bbox{r};E)$ will be the sum of all closed paths
traversed in the opposite direction. Averaging over disorder 
results in the vanishing of all contributions apart from those corresponding
to pairs of coherent paths. This coherence arises when the pair of paths 
involves scattering off the same impurities although not necessarily in 
the same order. Pictorially,  scattering off the same impurities  
is shown by impurity lines between $G^R$ and $G^A$. 
In the lowest perturbative order, the closed paths for $G^R$ and $G^A$ are
identical, apart from their different starting points $\bf r$
and $\bf r'$, while the impurity lines form a ladder
 which corresponds to a diffusion propagator
\begin{eqnarray}
\label{Diff}
P(q,\omega)=\frac{1}{2\pi\nu\tau^2}\frac{1}{Dq^2-i\omega}\,,
\qquad\quad
q\ell\ll 1, \;\omega\tau\ll 1
\end{eqnarray}
or its time-reversed counterpart, a Cooperon.

The key feature of the diagram for $R(\omega)$ generated in such a way,
Fig.~4, is that the starting points of the two
\ifpreprintsty \else
\begin{figure}
\vspace{0.3cm}
\hspace{0.02\hsize}
\epsfxsize=0.9\hsize
\epsffile{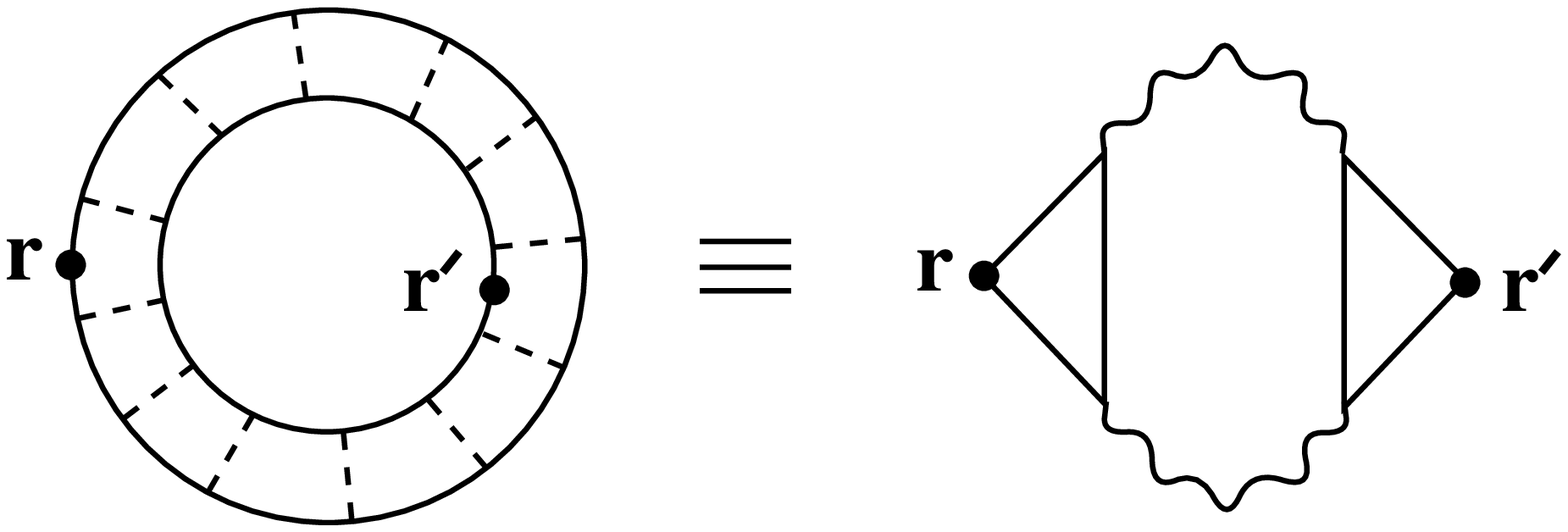}
\vspace{0.3cm}
 
\refstepcounter{figure}
\label{fig:4}\vspace{1cm}

{\small \setlength{\baselineskip}{10pt} FIG.\ \ref{fig:4}.
 The field theory diagram for the lowest order contribution 
to the two-level correlation function $R(\omega)$ in the standard
approach. This is identical to Fig. 1 except that the starting points
of the two trajectories, ${\bf r}$ and ${\bf r'}$ are distinguished.
In general inserting points in such a way yields many more diagrams
for $R(\omega)$ than for $F(\omega)$.
 }
\end{figure}
\fi
\noindent
 paths, $\bf r$ and $\bf r'$,
 are fixed in the beginning, even though we finally integrate over all
$\bf r$ and $\bf r'$ in Eq.\ (\ref{Reqn}).
 Contrarily, there are no starting points on the semiclassical diagrams
of Fig.~1. Obviously, one can
obtain the diagram of Fig.~4 from the semiclassical picture of Fig.~1
by inserting
points $\bf r$ and $\bf r'$ into different electron lines in all possible ways.
Inserting an external point is equivalent to taking a derivative
with respect to energy: inserting $\bf r$ replaces $G^R(E_1)$ by
$G^R(E_1)G^R(E_1)$, which can be achieved by the action of 
$-\partial/\partial E_1$. Similarly, inserting $\bf r'$ gives the same result
as $-\partial/\partial E_2$.  After averaging $R(\omega)$ depends
only upon $\omega=E_1-E_2$. Therefore, one may represent it as the second
derivative of a certain function $F(\omega)$,
\begin{eqnarray}
\label{RF}
R(\omega)=-\Delta^2\frac{\partial^2}{\partial\omega^2}F(\omega)\,.
\end{eqnarray}
We will show below that this $F(\omega)$ corresponds to  the `free energy'
of the appropriate field-theoretical functional.
Note that such a representation of $R(\omega)$
has previously been used in the evaluation of first-order
diagrams in the ballistic regime\cite{AG:95}. Here we show that
Eq.~(\ref{RF}) is valid to all orders of perturbation theory.

Higher order perturbative contributions to $R(\omega)$ arise from 
two paths where scattering occurs from
 the same impurities but in a different order. 
The parts of the diagrams
where the sequence of impurity scatterings
coincides  (or is time reversed) for the two paths
are again represented by diffuson (or Cooperon) ladders.
These ladders 
are connected by Hikami boxes which represent the change in the 
scattering sequence. In terms of paths, this corresponds to the 
regions of self-intersection described above.  The only difference
between these higher order contributions to $R(\omega)$ and semiclassical
diagrams in Figs.~2 and 3 is  the necessity to distinguish starting 
points $\bf r$ and $\bf r'$. Therefore these semiclassical diagrams describe 
 higher order contributions to  $F(\omega)$, and  $R(\omega)$ can be 
obtained with the help of Eq.~(\ref{RF}).

Finally let us further justify the above pictorial discussion with a rigorous
derivation using the standard field-theory 
machinery\cite{Weg:79,EfLKh,Hik:81,AKL:91}. We perform the
average over disorder by the replica trick. This can be done with either
bosonic (commuting) or fermionic (anticommuting) variables, and both must
yield the same results for physical quantities. In this
paper we will use bosonic variables. The  two-level correlation
function is then given by
 \ifpreprintsty \else\end{multicols} 
                  	\vspace*{-3.5ex} \hspace*{0.491\hsize} 
			\begin{minipage}{0.48\hsize}$\,$\hrule
			 \end{minipage}
\fi
\begin{eqnarray}
\label{Rnlsm}
R(\omega)=-\lim_{N\to0}\frac{\Delta^2}{16\pi^2N^2} \int {\cal D}Q
\left(\int \text{d}^dr \hbox{Tr}[\Lambda Q(\bbox{r})]\right)^2 
\exp{(-{\cal F}[Q;\omega])}-1\,,
\end{eqnarray}
 	\ifpreprintsty \else 
                  	\vspace*{-3.5ex} 
                  \begin{multicols}{2}
                   \fi\noindent
where ${\cal F}[Q;\omega]$ is the non-linear sigma model functional
\begin{eqnarray}
\label{nlsm}
{\cal F}[Q;\omega]=\frac{\pi\nu}{8}\int \text{d}^dr 
\hbox{Tr}\left[D(\nabla Q)^2-2i\omega\Lambda Q\right]
\end{eqnarray}
The matrix $Q$ has  rank $2N$:
we made $N$ replicas each for retarded and advanced Green's
functions. It satisfies the standard saddle-point
conditions:
$$Q^2=I\,,\qquad\quad \hbox{Tr}(Q)=0\,.
$$ 
The matrix $\Lambda$ is diagonal with
elements $+1$ for retarded indices and $-1$ for advanced indices. Then it
follows that 
$$\hbox{Tr}(\Lambda Q(\bbox{r}))=N(G^R-G^A)=-2\pi i N\nu(r)\,,$$ 
so that the
prefactor in Eq.\ (\ref{Rnlsm}) is the product of two densities of states.
If we now introduce the free energy $F(\omega)$ as
\begin{eqnarray}
\label{Fdef}
F(\omega)=\lim_{N\rightarrow 0}\frac{1}{N^2}
\int {\cal D}Q \exp{(-{\cal F}[Q;\omega])}\,,
\end{eqnarray}
we see
from Eqs.\ (\ref{Rnlsm}) and (\ref{nlsm}) that $R(\omega)$ can indeed be  
 written in terms of $F(\omega)$ as in Eq.\ (\ref{RF}).

 We can now apply the usual perturbative methods to
expand $F(\omega)$ in powers of the coupling constant (inverse
dimensionless conductance)
$1/g\equiv1/4\pi^2\nu D$ and then apply the relation
(\ref{RF}) to obtain $R(\omega)$. The great advantage of this
method is the much smaller number of diagrams we need to evaluate.
For example, in three-loop perturbation theory for the orthogonal case
there are 5 diagrams for $F(\omega)$ compared to $41$ for $R(\omega)$.

In previous work\cite{AKL:91} 
a source field for $\nu(E)$
has been introduced with a complex index structure similar to that 
necessary for calculating the conductance moments.
A considerable simplification here is that $\omega$ itself
suffices as a source field for calculating $\nu(E)$.
We note that this method cannot be extended to the supersymmetric
sigma model\cite{Ef:83} 
where the free energy is always unity owing to the fact
that the supersymmetry is preserved in the effective functional. To
break the supersymmetry one must introduce the $k$-matrix in the expression
for the   prefactors, $\hbox{STr}(k\Lambda Q)$, 
or equivalently introduce an appropriate
source term in the effective functional\cite{A2}. Although this is
straightforward, all `savings' in terms of the number 
of diagrams would be lost. Thus, for perturbative calculations the replicated
$\sigma$ model turns out to be more economical than the supersymmetric one. 
 Finally we note
that the free energy has previously been calculated up to four-loop
order\cite{Weg:87} in order to find $\beta(g)$. Our result above gives a
direct physical meaning to the field-theoretic free energy.

To generate perturbation theory we must introduce a parameterization of
$Q$ that satisfies the saddle-point constraints. There are several
parameterizations available, and they will give different contributions
for a given diagram, but the sum of all contributions at a given order
of perturbation theory will always be the same. The particular one used
depends upon the application. The choice
\begin{eqnarray}
\label{param1}
Q=\Lambda(W+\sqrt{1+W^2})\qquad
W=\left(\matrix{0&2iV\cr 2iV^+&0\cr}\right)
\end{eqnarray}
reproduces the results of the original impurity diagrammatics in the
diffusive regime\cite{Hik:81}, 
and is thus useful for direct comparison between the
results of impurity diagrammatics and the non-linear $\sigma$-model.
Here $V$ is an unconstrained $N\times N$ matrix with elements of the
form appropriate for the given Dyson ensemble. The functional integration
is then carried out over the independent variables $V$.
Another parameterization,\cite{AKL:91}
\begin{eqnarray}
\label{param2}
Q=\Lambda(1+W/2)(1-W/2)^{-1}
\end{eqnarray}
where $W$ has the same properties as before, is more convenient for
calculation because many terms then vanish, and the Jacobian of the
transition from $Q$ to $W$ is unity in the replica limit. Since the
Jacobian contributes terms necessary
to remove ultraviolet divergences, this means
that the sum of all terms at any order in perturbation theory will have
no ultraviolet divergence and so there is no need for regularization. 
Considerations in the ergodic regime are also much simpler, since Jacobian 
terms would contribute there.
In what follows we will, therefore, use these parameterization.

When we substitute a parameterization into the free energy functional
${\cal F}[Q]$ we obtain a sum of vertices,
\begin{eqnarray}
\label{Fseries}
{\cal F}[Q]=\sum_{n=1}^\infty {\cal F}_{2n}[W]
\end{eqnarray}
where the ${\cal F}_{2n}[W]$ vertex contains $2n$ powers of $W$. The first
term ${\cal F}_2$ is the quadratic part, and leads to the ladder propagators.
The perturbation expansion up to the third loop order
is then obtained by expanding out the
exponential in the other ${\cal F}_{2n}$,
\begin{eqnarray*}
\exp{\left(-{\cal F}_4-{\cal F}_6-{\cal F}_8-\dots\right)}=
1-{\cal F}_4-\left({\cal F}_6-{\cal F}_4^2/2\right)
\end{eqnarray*}
and averaging with respect to $\exp{(-{\cal F}_2)}$. Upon substituting the
parameterization we get
\ifpreprintsty \else\end{multicols} 
                  	\vspace*{-3.5ex} \hspace*{0.491\hsize} 
			\begin{minipage}{0.48\hsize}$\,$\hrule
			 \end{minipage}
\fi
\begin{eqnarray}
\nonumber {\cal F}_2&=&\pi\nu\int d^dr\hbox{Tr}\left[-D\nabla V\nabla V^+ 
-i\omega VV^+\right]\\
{\cal F}_4&=&\pi\nu\int d^dr\hbox{Tr}\left[-2D\nabla V\nabla V^+ VV^+
\label{Fexp}
-i\omega VV^+VV^+\right]\\
\nonumber
{\cal F}_6&=&\pi\nu\int d^dr\hbox{Tr}\left[D(2\nabla V\nabla V^+ VV^+
+\nabla VV^+V\nabla V^+ VV^+)-i\omega VV^+VV^+VV^+\right]
\end{eqnarray}
 	\ifpreprintsty \else 
                  	\vspace*{-3.5ex} 
                  \begin{multicols}{2}
                   \fi\noindent
If we then Fourier transform into momentum space we recover the
expected diffusion propagator of Eq.\ (\ref{Diff}) as the average of
$VV^+$ with respect to $\exp{(-{\cal F}_2)}$,
\begin{eqnarray}
\label{vvav}
\langle V_{i\alpha}(q)V^+_{\beta j}(-q)\rangle =
\frac{1}{2\pi\nu}\frac{1}{Dq^2-i\omega}\delta_{ij}\delta_{\alpha\beta}
\end{eqnarray}
The higher-order terms yield the effective four- and six-point vertices
\begin{eqnarray}
\label{Hikami}
V_4(q_i)&=&2\pi\nu\left[-D\Delta_4^1-2i\omega\right]\\
V_6(q_i)&=&-2\pi\nu\left[-D(\Delta_6^1+\Delta_6^3)-3i\omega\right]
\end{eqnarray}
where $\Delta_n^m$ is the sum of all distinct scalar products of
pairs $q_i$ and $q_{i+m}$ of the $n$ incoming momenta  which are 
separated by $m$ vertices,
\begin{eqnarray}
\label{Delta}
\Delta_n^m=\sum q_i.q_{i+m}
\end{eqnarray}
Note that in the derivation of Eq.\ (\ref{Hikami}) we have symmetrized
over all incoming momentum variables.

If we use the above formulas to calculate the expansion of free energy
$F(\omega)$ to three-loop
order, the diagrams obtained are simply those we derived earlier using
the trajectory language. The field theory enables us to associate an
algebraic expression to each component in a diagram, and we obtain the
results
\begin{eqnarray}
F_1&=-&\sum_{q_1}\ln{{\cal P}_1}\nonumber\\
F_2&=&\frac{1}{(2\pi\nu)}\sum_{q_1,q_2}\frac{i\omega}{{\cal P}_1{\cal P}_2}\nonumber\\
F_{3a}&=&\frac{2}{(2\pi\nu)^2}\sum_{q_1,q_2,q_3}
\frac{(i\omega)^2}{{\cal P}_1^2{\cal P}_2{\cal P}_3}\nonumber\\
\label{3loop}
F_{3b}&=&-\frac{1}{(2\pi\nu)^2}\sum_{q_1,q_2,q_3}
\frac{{\cal P}_1}{{\cal P}_1{\cal P}_2{\cal P}_3}\\
F_{3c}&=&-\frac{1}{(2\pi\nu)^2}\sum_{q_1,q_2,q_3}
\frac{{\cal P}_1-2i\omega}{{\cal P}_1{\cal P}_2{\cal P}_3}\nonumber\\
F_{3d}&=&\frac{1}{4(2\pi\nu)^2}\sum_{q_1,q_2,q_3}
\frac{({\cal P}_{12}-i\omega)^2}{{\cal P}_1{\cal P}_2{\cal P}_3{\cal P}_{123}}\nonumber\\
F_{3e}&=&\frac{1}{2(2\pi\nu)^2}\sum_{q_1,q_2,q_3}
\frac{({\cal P}_{12}-i\omega)({\cal P}_{13}-i\omega)}{{\cal P}_1{\cal P}_2{\cal P}_3{\cal P}_{123}}
\nonumber
\end{eqnarray}
where we use the notation 
$$
{\cal P}_{1\dots n}=D(q_1+\dots q_n)^2-i\omega\,.$$ Of the above terms only
$F_1$, $F_{3b}$ and $F_{3d}$ contribute in the unitary case. All
diagrams contribute in the orthogonal case, and there is an extra factor
of two arising from the two possible relative directions of the electron
lines.

Symbolic expressions for $F$ similar to those in Eqs.\ (\ref{3loop}) have been 
previously derived in Ref.\ \onlinecite{CLS2} where the parameterization
(\ref{param1}) has been used. 
Equations  (\ref{3loop}) derived within the parameterization (\ref{param2}) 
are valid for to both the ergodic and diffusive regimes.

Let us first check that  these formulae
reproduce the known results for the ergodic regime to this order.
The ergodic regime corresponds to the zero-mode approximation in which
we set all $q_i$ equal to zero. Doing this and differentiating twice
with respect to $\omega$ yields
\begin{mathletters}
\begin{eqnarray}
R^{\text{ort}}(x)&=&\Re \text{e}\left[\frac{1}{(ix)^2}+\frac{1}{(ix)^3}
+\frac{3}{2(ix)^4}\right]\\
R^{\text{uni}}(x)&=&\Re \text{e}\left[\frac{1}{2(ix)^2}\right]
\label{Rou}
\end{eqnarray}
\end{mathletters}
where $x=\pi\omega/\Delta$. These are the correct expansions
of the exact results.\cite{Ef:83} Indeed, for $K(t)$, which is
the Fourier transform of 
$R(x)$, one has \cite{Ef:83} 
 in the region $0\le t\ll 1$:
\begin{mathletters}
\begin{eqnarray}
\label{ergo1}
K^{\text{ort}}(t)&=&t[2-\ln(t+1)]\approx 2t-t^2+\frac{1}{2}t^3\\
K^{\text{uni}}(t)&=&t\theta(1-t)=t
\end{eqnarray}
\end{mathletters}
Here $t$ is measured in units of $t_H=\hbar/\Delta$.
(To compare the results of Eqs.\ (\ref{Rou}) and  (\ref{ergo1}), one uses
the result that the inverse Fourier transform of $t^n$ is $i^{n+1}n!/2x^{n+1}$.)
Recall the famous result that there are no corrections to the universal
RMT result in the unitary case to any order in perturbation theory.
This happens due to the cancellation between rather than the absence
of appropriate diagrammatic contributions, and is therefore a useful
check on our calculation. Note also that the complete
perturbative expansion for $R(\omega)$
would give, upon Fourier transformation, the
exact $K(t)$ only for $t<1$. The perturbation theory cannot give the
discontinuity point, $t=1$,  because this is controlled by the non-trivial
saddle point\cite{A2}.

We will now derive the leading-order contributions to $R(\omega)$ in
2d.  This was previously done\cite{KL:95a} for the orthogonal case, 
and the form in the unitary case conjectured from a one-parameter
scaling hypothesis. We replace the sums over $q$ in Eq.\
by integrals and use dimensional regularization to evaluate
these integrals in $d=2+\epsilon$ dimensions. (Note that all the relevant
integrals may be found in Ref.\ \onlinecite{Weg:87}). Then we carefully take
the limit $\epsilon\rightarrow 0$, keeping only  the terms divergent
in this limit. The two-loop orthogonal result is
\begin{eqnarray*}
F^{\text{ort}}_{2}(\omega)=\frac{\Delta^2L^d}{\pi^2\beta}
\frac{1}{(2\pi\nu)(2\pi D)^2}
\frac{(-i\omega)^{\epsilon+1}}{\epsilon^2}
\end{eqnarray*}
Taking the second derivative with respect to $\omega$, and noting
that dimensionless conductance $g=G/(e^2/\pi h)=4\pi^2\nu D$, gives
\begin{eqnarray}
\label{Ro2d}
R^{\text{ort}}_{2}(\omega)=\frac{2\Delta}{\pi g^2}\Re \text{e}\left[
\frac{(-i\omega)^{\epsilon-1}}{\epsilon}\right]
\end{eqnarray}
Finally we let $\epsilon\rightarrow 0$ and note that 
$(-i\omega)^{\epsilon}/\epsilon$ becomes $\ln{(-i\omega\tau)}$.
Hence
\begin{eqnarray*}
R^{\text{ort}}_{2}(\omega)=\frac{2\Delta}{\pi g^2}\Re \text{e}\left[
\frac{\ln{(-i\omega\tau)}}{i\omega}\right]=
-\frac{1}{g^2s}\,,\qquad s\equiv\frac\omega\Delta
\end{eqnarray*}
The symplectic result would simply follow upon multiplying by $-1/2$.

Next we consider the three-loop unitary case. Careful evaluation of the
integrals yields
\begin{eqnarray*}
F^{\text{uni}}_{3}(\omega)=\frac{\Delta^2L^d}{\pi^2\beta}
\frac{1}{(2\pi\nu)^2(2\pi D)^3}
\frac{(-i\omega)^{3\epsilon/2+1}}{3\epsilon^2}\,.
\end{eqnarray*}
Thus 
\begin{eqnarray}
\label{Ru3d}
R^{\text{uni}}_{3}(\omega)=\frac{\Delta}{\pi g^3}\Re \text{e}\left[
\frac{(-i\omega)^{3\epsilon/2-1}}{2\epsilon}\right]
\end{eqnarray}
which in the limit $\epsilon\rightarrow 0$ becomes
\begin{eqnarray}
\label{Ru3dres}
R^{\text{uni}}_{3}(\omega)=\frac{3\Delta}{4\pi g^3}\Re \text{e}\left[
\frac{\ln{(-i\omega\tau)}}{i\omega}\right]=
-\frac{3}{8g^3s}\,.
\end{eqnarray}

Finally we calculate the three-loop orthogonal result:
\begin{eqnarray*}
F^{\text{ort}}_{3}(\omega)=\frac{\Delta^2L^d}{\pi^2\beta}
\frac{1}{(2\pi\nu)^2(2\pi D)^3}
\frac{(4-3\epsilon)(-i\omega)^{3\epsilon/2+1}}{3\epsilon^3}\,,
\end{eqnarray*}
so that
\begin{eqnarray}
\label{Ro3d}
R^{\text{ort}}_{3}(\omega)=\frac{2\Delta}{\pi g^3}
\frac{(4+3\epsilon)(-i\omega)^{3\epsilon/2-1}}{2\epsilon^2}\,,
\end{eqnarray}
which in the limit $\epsilon\rightarrow 0$ gives
 \ifpreprintsty \else\end{multicols} 
                  	\vspace*{-3.5ex} \hspace*{0.491\hsize} 
			\begin{minipage}{0.48\hsize}$\,$\hrule
			 \end{minipage}
\fi
\begin{eqnarray*}
R^{\text{ort}}_3(\omega)=\frac{9\Delta}{2\pi g^3}\Re \text{e}\left[
\frac{\ln^2{(-i\omega\tau)}+\ln{(-i\omega\tau)}}{i\omega}\right]
=\frac{9}{4g^3s}\Bigl[1+2\ln({s\Delta\tau})\Bigr]\,.
\end{eqnarray*}

We now show that the same results can be obtained using a one-parameter
scaling hypothesis in which the renormalized conductance, $g(\omega)$,
is substituted into the lowest order $R(\omega)$ diagram. The latter has the
form
$R(\omega)=({\Delta^2}/{\pi^2\beta})\Re \text{e}
\sum_q{\cal P}_1^{-1}$, 
and substituting here $D=D_0+\delta D(\omega)$ gives upon the expansion 
in $\delta D$
\begin{eqnarray}
\label{RGform}
R(\omega)=\frac{\Delta^2}{\pi^2\beta}\left\{
\sum_q\frac{1}{(D_0q^2-i\omega)^2}-
2\left(\frac{\delta D}{D_0}\right)
\sum_q\frac{D_0q^2}{(D_0q^2-i\omega)^3}+
3\left(\frac{\delta D}{D_0}\right)^2
\sum_q\frac{(D_0q^2)^2}{(D_0q^2-i\omega)^4}+\dots\right\}
\end{eqnarray}
	\ifpreprintsty \else 
                  	\vspace*{-3.5ex} 
			\begin{minipage}{0.48\hsize}$\,$\hrule
			 \end{minipage}\vspace*{1.5ex}
                  \begin{multicols}{2}
                   \fi
We obtain the perturbative correction to the diffusion constant,
$\delta D(\omega)$, from the $\beta(g)$ function which is defined by
\begin{eqnarray*}
\beta(g)=\frac{d\ln{g(L)}}{d\ln{L}}
\end{eqnarray*}
To two-loop order the orthogonal and unitary beta functions are
$\beta^{\text{ort}}(g)={2}/{g}$ and
$\beta^{\text{uni}}(g)={2}/{g^2}$
from which it follows that
\begin{eqnarray*}
\frac{\delta D_{\text{ort}}}{D_0}=
\frac{\delta g_{\text{ort}}}{g_0}=-\frac{2}{g_0}\ln{(L/\ell)}\to
\frac{2}{g_0}\frac{(-i\omega)^{\epsilon/2}}{\epsilon}\\
\frac{\delta D_{\text{uni}}}{D_0}=
\frac{\delta g_{\text{uni}}}{g_0}=-\frac{2}{g_0^2}\ln{(L/\ell)}\to
\frac{1}{g_0^2}\frac{(-i\omega)^\epsilon}{\epsilon}
\end{eqnarray*}
We can then substitute these results into Eq.\ (\ref{RGform}), and
use the following results for the $q$-integrals,
\begin{eqnarray*}
\sum_q\frac{D_0q^2}{(D_0q^2-i\omega)^3}=-\frac{2+\epsilon}{8\pi D}
(-i\omega)^{\epsilon/2-1}\\
\sum_q\frac{(D_0q^2)^2}{(D_0q^2-i\omega)^4}=
-\frac{4+3\epsilon}{16\pi D}(-i\omega)^{\epsilon/2-1}
\end{eqnarray*}
We find that this exactly reproduces the results of 
Eqs.\ (\ref{Ro2d}), (\ref{Ro3d}) and (\ref{Ru3d}), proving the
validity of the scaling hypothesis to three-loop order. Note that the
conjecture of Ref.~\onlinecite{KL:95a} for $R_3^{\text{uni}}(\omega)$ does not   produce
the correct numerical factor in Eq.\ (\ref{Ru3dres}). The reason is that 
the calculation in  Ref.\ \onlinecite{KL:95a} has been performed in exactly two
dimensions where it was possible to take into account all weak-localization
logarithms but not logarithmic corrections arising from stronger 
mesoscopic divergences. To pick up all logarithmic corrections one must
work in $d=2+\epsilon$ and take carefully 
the limit $\epsilon\rightarrow 0$, as we did here.

In conclusion, we have shown that the language of semiclassical trajectories
suggests the most economical way of drawing diagrammatic corrections to
spectral correlations in disordered electronic systems
up to a high order of perturbation theory. 
We have introduced the weak diagonal approximation which includes the contributions
of pairs of trajectories which are made from identical pieces joined together
in different ways at some self-intersection (in real space) points. This gives
physical meaning to the loop expansion of the field-theoretical `free energy':
the second derivative of this free energy with respect to frequency is the 
two-level correlation function. We have shown this directly by using the 
replicated nonlinear $\sigma$ model. Note that such a derivation does not
work for the supersymmetric  nonlinear $\sigma$ model as the free energy
equals zero unless the supersymmetry is broken. Naturally, all the perturbative
results may be reproduced within  the supersymmetric model but in a much less
economical way.
Using the method described above, we have calculated the leading
order contributions to the two-level correlation function in $2d$ in the 
non-ergodic regime where the standard diagonal
approximation gives vanishing results. 

In the ergodic regime we have found that 
the loop expansion reproduces the deviation from  
linear in time behavior of the spectral form factor in the orthogonal case known
from random matrix theory. In the unitary case, we have demonstrated that all 
such corrections are mutually cancelled. The random matrix theory is believed
to describe correctly level statistics of classically chaotic systems 
in the universal regime.$^{5-7}$ It is therefore reasonable to conjecture 
that the weak diagonal approximation introduced here should be valid for
generic chaotic systems. Although the universal regime is well understood
within the supersymmetric approach, the weak diagonal approximation  might be 
extended to the non-ergodic regime in chaotic systems. 

\acknowledgements
 Work in Birmingham was supported by EPSRC grant
GR/J35238.
I.V.L.\ and B.L.A.\ thank the  ITP in
Santa Barbara for  kind hospitality and partial support under NSF 
Grant No.\ PHY94-0719.
R.A.S.\ and I.V.L.\  thank the Newton Institute at Cambridge
for  kind hospitality at the final stage of this work.

\ifpreprintsty\newpage

\section{Figure Captions}
\medskip
FIG. (1). (a) The 1-loop diagram for the free energy  $F(\omega)$, 
which consists of pairs of trajectories that are identical up to 
time reversal symmetry. In the field theory language it consists of 
a single closed wavy line. 
(b) The same diagram in the disordered metal. Since the closed wavy
line in (a) consists of a sum over impurity ladders, and the diagram
with $n$ impurity ladders has symmetry factor $1/n$, the ladder gives
logarithmic contribution $-\ln{(Dq^2-i\omega)}$.
\medskip
\par\noindent
FIG. (2). The 2-loop diagram for the  free energy $F(\omega)$ 
corresponding to paths with one self-intersection. First we show 
the two different ways of linking up the paths at the 
self-intersection point. Then we rewrite this in the field theory 
language so that regions where the two paths are identical become 
wavy lines, whilst the region of self-intersection becomes a closed 
box. Finally we twist the latter diagram into its usual form.
\medskip
\par\noindent
FIG. (3). The five 3-loop diagrams for the free energy $F(\omega)$ 
corresponding to paths with two self-intersections. We follow the 
format of Fig. 2, showing first the pairs of distinct trajectories; 
then rewriting these with wavy lines representing regions where the 
two paths are identical, and closed boxes at the self-intersection 
points; then finally twisting the latter diagrams in their usual form.
\medskip
\par\noindent
FIG. (4). The field theory diagram for the lowest order contribution 
to the two-level correlation function $R(\omega)$ in the standard
approach. This is identical to Fig. 1 except that the starting points
of the two trajectories, ${\bf r}$ and ${\bf r'}$ are distinguished.
In general inserting points in such a way yields many more diagrams
for $R(\omega)$ than for $F(\omega)$.

\newpage
\vspace{0.3cm}
\hspace{0.02\hsize}
\epsfxsize=0.9\hsize
\epsffile{sla1.ps}

\vfill

\centerline{\large Fig.~1}

\newpage
\vspace{0.3cm}
\hspace{0.02\hsize}
\epsfxsize=0.9\hsize
\epsffile{sla2.ps}

\vfill
 
\centerline{\large Fig.~2}

\newpage
\vspace{0.3cm}
\hspace{0.02\hsize}
\epsfxsize=0.9\hsize
\epsffile{sla3.ps}
 
\vfill
 
\centerline{\large Fig.~3}

\newpage
\vspace{0.3cm}
\hspace{0.02\hsize}
\epsfxsize=0.9\hsize
\epsffile{sla4.ps}
 
\vfill
 
\centerline{\large Fig.~4}

\else\end{multicols}\fi\end{document}